\documentclass[manuscript, nonacm]{acmart}
\usepackage{hyperref} 
\usepackage{float}
\usepackage{booktabs}
\usepackage{longtable}
\usepackage{multirow}
\usepackage{balance}
\usepackage{bibentry}
\usepackage{makecell}
\usepackage{soul}
\usepackage{pdflscape}

\usepackage{xcolor} 

\usepackage{listings}

\lstdefinestyle{studentsnapshotlisting}{
    frame=tb,
    language=python,
    aboveskip=3mm,
    belowskip=3mm,
    showstringspaces=false,
    columns=flexible,
    basicstyle={\small\ttfamily},
    numbers=left,
    numberstyle=\color{gray},
    keywordstyle=\color{blue},
    commentstyle=\color{dkgreen},
    stringstyle=\color{mauve},
    breaklines=true,
    breakatwhitespace=true,
    tabsize=3,
    upquote=true,
    escapeinside=||
}

\lstdefinestyle{tablelisting}{
    frame=tb,
    language=python,
    frame=none,
    showstringspaces=false,
    columns=flexible,
    basicstyle={\small\ttfamily},
    keywordstyle=\color{blue},
    commentstyle=\color{dkgreen},
    stringstyle=\color{mauve},
    breaklines=true,
    breakatwhitespace=true,
    breakautoindent=false,
    breakindent=0pt,
    tabsize=3,
    upquote=true,
}

\let\origthelstnumber\thelstnumber
\makeatletter
\newcommand*\Suppressnumber{%
  \lst@AddToHook{OnNewLine}{%
    \let\thelstnumber\relax%
     \advance\c@lstnumber-\@ne\relax%
    }%
}

\newcommand*\Reactivatenumber{%
  \lst@AddToHook{OnNewLine}{%
   \let\thelstnumber\origthelstnumber%
   \advance\c@lstnumber\@ne\relax}%
}

\makeatother

\AtBeginDocument{%
  \providecommand\BibTeX{{%
    \normalfont B\kern-0.5em{\scshape i\kern-0.25em b}\kern-0.8em\TeX}}}

\definecolor{dkgreen}{rgb}{0,0.6,0}
\definecolor{gray}{rgb}{0.5,0.5,0.5}
\definecolor{mauve}{rgb}{0.58,0,0.82}

\begin{document}

\title{An Investigation Into Secondary School Students' Debugging Behaviour in Python}

\author{Laurie Gale}
\affiliation{%
    \department{Raspberry~Pi~Computing~Education Research Centre}
    \institution{University of Cambridge}
    \city{Cambridge}
    \country{UK}
}
\email{lpg28@cst.cam.ac.uk}

\author{Sue Sentance}
\affiliation{%
    \department{Raspberry~Pi~Computing~Education Research Centre}
    \institution{University of Cambridge}
    \city{Cambridge}
    \country{UK}
}
\email{ss2600@cst.cam.ac.uk}

\begin{abstract}
\textbf{Background and context:} Debugging is a significant and often frustrating challenge for beginner programmers. Understanding students' debugging behaviours and strategies can help to identify common difficulties and inform approaches for alleviating these. Currently, there are limited studies of school students’ debugging behaviour in a text-based programming language, a medium through which millions are learning to program.

\textbf{Objectives:} In this paper, we investigate the debugging behaviour of 12-14-year-old students learning Python through a lesson-long classroom study.

\textbf{Method:} We collected program snapshots from 73 students' attempts at a set of Python debugging exercises in an online code editor. Through qualitative content analysis of these snapshots, we developed a granular categorisation of the changes students made when debugging.

\textbf{Findings:} A range of debugging behaviours were exhibited by students, many of which were ineffective. Students added errors through small-scale changes, reverted corrective changes, and repeatedly ran identical programs in quick succession. From the results, we identify four barriers to successful and reliable debugging for students learning a text-based programming language: fragile knowledge, a lack of systematicity and reflection, the syntax barrier, and dynamics of emotions and attitudes.

\textbf{Implications:} This paper highlights some of the difficulties that secondary school students have when debugging in Python and the challenges of analysing program snapshots through manual inspection. We recommend that school teachers explicitly teach a systematic approach to debugging and discourage the use of ineffective debugging behaviours, and that programming environments should contain features that facilitate successful debugging.
\end{abstract}

\ccsdesc[500]{Social and professional topics~K-12 education}

\keywords{debugging, debugging behaviour,  programming education, computing education}

\maketitle

\section{Introduction}
Debugging, the process of finding and fixing errors in computer programs, is a vital skill for any programmer. Encountering errors cannot be avoided when programming; even professional software developers report spending a significant portion of their time debugging \citep{ReversibleDebuggingSoftware, StudyingAdvancementProfessionalDebuggingPractice}. For beginner programmers, debugging proficiency reduces some of the difficulties of learning to program, with some studies reporting an association between `good debuggers' and good programmers \cite{NottinghamDebuggingBehaviour, FindingFixingFlailing}.

The use of a text-based programming language (TBPL) in secondary school is mandated in several computing curricula (e.g., \citep{UKSecondaryComputingCurriculum, ScotlandTechnologiesCurriculum, Swedish2024Curriculum, BritishColumbiaADSTCurriculum, ProgrammingSkillsInLatinAmerica}) and is likely to become more prevalent in the future. While this exposes students to traditional programming languages, debugging is a common source of struggle and emotional discomfort for students. Some students have trouble locating errors \citep{ThinkAloudNoviceDebugging, NottinghamDebuggingBehaviour, AnalysingIntroductoryDebuggingProcess} and often use ineffective strategies which cause errors to persist in their programs (e.g., \citep{ConditionsLearningNovices, ThinkAloudNoviceDebugging, FindingFixingFlailing}). Struggling to resolve errors can be frustrating and anxiety-inducing \citep{EDAProgrammingEmotions, ProgrammingAssignmentsEmotionalToll,
AnxietyLearningToProgram}, which may be especially true for younger learners. If these struggles come during students' early experiences of learning to program, they will likely tarnish their attitudes towards programming \citep{CS1SelfEfficacyPerceptions, ProgrammingAssignmentsEmotionalToll, ProgrammingSelfEfficacySelfAssessment}.

Although some studies have investigated school students' debugging behaviour, most of them have not involved TBPLs \citep{InvestigatingDebuggingProcessesScopingReview}. These languages are generally more complex than block- or game-based counterparts, provide a larger scope for errors, and may afford different debugging strategies. Increasing our understanding of school students' debugging behaviours and strategies in these languages would help to inform pedagogical approaches for teaching debugging in the classroom.

This paper reports on a study of secondary school students' (aged 12-14 years old) debugging behaviour in Python. Students learning Python attempted to debug five erroneous Python programs using an online code editor in one of their computing lessons. Snapshots of students' programs were collected every time they were run, which we analysed with qualitative content analysis to answer the following research question:

\begin{itemize}
    \item What behaviours are exhibited by secondary school students when debugging erroneous text-based programs?
\end{itemize}

Based on our findings and discussion, this paper has the following contributions.

\begin{itemize}
    \item A granular categorisation of secondary school students' debugging behaviour in Python, reported in Section \ref{sec:results}.
    \item The identification of four barriers to successful and reliable debugging for students learning a text-based programming language (fragile knowledge of the programming constructs they are trying to fix, a lack of systematicity and reflection in their approach, the syntax barrier of the programming language, and dynamics of attitudes and emotions), discussed in Section \ref{sec:barriers-effective-debugging}.
    \item Recommendations for how school teachers could improve their students' debugging behaviour, discussed in Section \ref{sec:conclusions-further-work}.
\end{itemize}


\section{Beginner Programmers' Debugging Strategies}\label{sec:related-work}
Debugging is an iterative process that begins when encountering a failure in the program's execution, either through running or testing. Ideally, a programmer should then iteratively complete the following steps until the error is resolved: ensure they understand the problem space, identify a discrepancy between the program's actual and expected behaviour, isolate the location of an error(s), attempt to resolve the error(s), and test their fix(es) \citep{AssessingDebuggingFormalModel, TeachingDebuggingFramework, GeneralFrameworkDebugging}. For beginner programmers, a few instances of runtime and syntax errors\footnote{See Table \ref{tab:error-type-definitions} for definitions.} are often responsible for a large proportion of errors (e.g., \citep{HighSchoolIntroProgrammingErrors, 37MillionCompilations, AllSyntaxErrorsNotEqual, LargeScaleLogoErrors, ChineseMiddleSchoolDifficulties}). Particularly severe\footnote{The \textit{severity} of an error is the product of its frequency and the time taken to resolve it \citep{NewLookNoviceProgrammingErrors}.} semantic errors include incorrect variable naming or declaration \citep{ErrorMessageCausePython, ErrorLandscapeCharacterising, MeaningfulCategorisationNoviceErrors, HighSchoolIntroProgrammingErrors} and type errors \citep{37MillionCompilations, ErrorMessageCausePython, ErrorLandscapeCharacterising, AllSyntaxErrorsNotEqual}, with \citet{ErrorMessageCausePython} finding type and name errors accounting for nearly half of the errors made by high school students learning Python. While syntax errors may be easily resolvable `slips', such as missing semi-colons or unbalanced parentheses \citep{37MillionCompilations, ErrorMessageCausePython, WritingBetweenLines}, some studies have documented the `syntax barrier' that beginners face \citep{UnderstandingSyntaxBarrier, AllSyntaxErrorsNotEqual}. \citet{AllSyntaxErrorsNotEqual} report undergraduates taking several minutes to resolve the aforementioned slips, while some students may struggle to write any syntactically correct code \citep{UnderstandingSyntaxBarrier, JadudCompilationBehaviour}.

Upon encountering these errors, students employ some debugging \textit{strategies} to try and resolve them. \citeauthor{InteractionDomainStrategicKnowledge} define a strategy as a ``goal-directed procedure that is planfully or intentionally evoked either prior to, during, or after the performance of a task'' \citep[p. 376]{InteractionDomainStrategicKnowledge}. Strategies consist of lower-level, observable \textit{behaviours}, such as testing a program with an extreme value or undoing a previous change. The debugging strategies and behaviours that students employ have a significant impact on their success in resolving errors. These are important for teachers and researchers to understand, so that useful feedback can be provided and teaching approaches that improve students' strategies can be developed.

Researchers have investigated students’ debugging strategies and behaviours using a range of methods, with notably few focused on school students learning a TBPL. In \citeauthor{InvestigatingDebuggingProcessesScopingReview}'s review of students' debugging processes, only six of the 36 studies they surveyed focused on this context \citep{InvestigatingDebuggingProcessesScopingReview}. In an early observational study of school students' behaviour in LOGO and BASIC, \citet{ConditionsLearningNovices} identified two behavioural profiles related to debugging; \textit{stoppers}, who abandon any effort upon encountering a difficulty, and \textit{movers}, who repeatedly change and run their code without stopping long enough to appear stuck. Two subtypes of mover are \textit{extreme movers}, who move so quickly that it is not possible to make intentional changes, and \textit{tinkerers}, who repeatedly make small changes to their program in an attempt to get it working. Where the speed of their changes characterises extreme movers, tinkerers are characterised by the magnitude of them. We consider tinkering synonymous with trial and error.

Similar studies of school students have commonly reported ineffective debugging behaviours. \citet{ProblemSolvingGroupsLearningProgramming} observed a common tendency among grade 6-8 students using BASIC to `opportunistically' debug by attempting fixes soon after an error was encountered. Similarly, the majority of grade 12 students surveyed by \citet{NoviceDebuggingVBProgramming} reported modifying erroneous VB programs without any debugging plan. Students have also been observed tinkering, undoing corrective changes, and deleting large portions of code in game- and block-based environments \citep{ProblemSolvingDebuggingK68, ElementaryGameBasedDebugging, ElementaryPuzzleBasedDebugging}. \citet{ProblemSolvingDebuggingK68} believed these behaviours were due to a lack of initial program comprehension, which hindered students' ability to localise errors. On the other hand, students who were more successful at debugging in \citet{MAADSMethod} were characterised by small, targeted changes, which likely relates to their ability to self-regulate \citep{StudentsDebuggingRegulationProcessComputationalModelling}. A commonly observed behaviour in line with stopping is the repeated running of identical erroneous programs that do not compile \citep{MAADSMethod, ProblemSolvingDebuggingK68, MeaningfulCategorisationNoviceErrors}. This behaviour is of no benefit for resolving errors and may indicate disengagement with the debugging process \citep{ConditionsLearningNovices}.

In classroom settings, students may also collaborate. Students who debug with peers likely influence each other's behaviour \citep{ProblemSolvingGroupsLearningProgramming}, engage in constructive dialogue \citep{PairDebuggingTransactiveDiscourseAnalysis}, and regulate each other's learning \citep{ExploringCollabDebuggingProcesses}. Alternatively, students may seek the support of their teacher, who can encourage them to articulate their thoughts about an error \citep{SupportingStudentsScienceInquiry, ReflectionsSustainedDebuggingSupport, TeacherSupportForDebuggingPhysicalComputing}, support them through the debugging process \citep{ReflectionsSustainedDebuggingSupport, TeacherDiagnosticInterventionSkills}, or enable the use of more sophisticated strategies \citep{DebuggingPathways}. However, given that some school teachers have reported an overwhelming reliance on them for debugging support \citep{CurrentPerspectivesDebugging, HandsUpProblem}, such beneficial teacher-student interactions may be rushed or simply not possible.

A range of behaviours has also been found among studies of CS1 students debugging in a TBPL. Reflective behaviours, such as tracing and monitoring of program state through print statements, are more commonly observed compared to studies with school students \citep{ThinkAloudNoviceDebugging, ComparisonDebuggingBehaviourNoviceExpert, NottinghamDebuggingBehaviour}. However, behaviours in line with tinkering and extreme moving have also been reported \citep{ThinkAloudNoviceDebugging, ProgrammingAssignmentsEmotionalToll}, even when students' programs are syntactically incorrect \citep{JadudCompilationBehaviour, UnderstandingSyntaxBarrier}. In practice, this may involve students making speculative changes to parts of a program they do not understand and may be more common among less successful students \citep{ThinkAloudNoviceDebugging}. When any ineffective approaches are taken to fix errors, changes are both unlikely to be corrective and likely to add further errors to the program \citep{ConditionsLearningNovices, ComparisonDebuggingBehaviourNoviceExpert, MAADSMethod, ThinkAloudNoviceDebugging, DebuggingNoviceSkilledProgrammers}.

These often ineffective behaviours differ from experienced programmers' debugging practice, who use specialist debugging tools such as step-by-step execution and variable inspection to isolate errors \citep{StudyingAdvancementProfessionalDebuggingPractice, HowProgrammersDebugTheirCode, DichotomyDebuggingBehaviour}. While these tools are too complex for many beginners \citep{ReviewGenericPVSs, StudentBarriersInfluencingDebuggerUsage}, experienced programmers have more expertise in several other areas, including their ability to comprehend code and manage hypotheses about the cause of an error(s) \citep{ComparisonDebuggingBehaviourNoviceExpert, DebuggingNoviceSkilledProgrammers}. In contrast, beginners' programming knowledge is often \textit{fragile}, which \citeauthor{FragileKnowledgeOriginalPaper} define as ``partial, hard to access, and often misused'' \citep[p. 4]{FragileKnowledgeOriginalPaper}. Fragile knowledge may not always prevent students from successfully resolving errors, but does prevent them from consistently doing so \citep{NottinghamDebuggingBehaviour, ThinkAloudNoviceDebugging}. When students are unaware of how to resolve an error, they may feel they have little choice but to tinker.

In summary, beginner programmers employ a range of strategies and behaviours when debugging their programs. While some of these facilitate consistently successful debugging \citep{MAADSMethod, ThinkAloudNoviceDebugging, ComparisonDebuggingBehaviourNoviceExpert}, tinkering, extreme moving, and stopping have been frequently observed in studies with school students \citep{ProblemSolvingDebuggingK68, ElementaryPuzzleBasedDebugging, ElementaryGameBasedDebugging, ConditionsLearningNovices}. While tinkering can be a positive strategy \citep{ConditionsLearningNovices}, doing so unsystematically can easily accumulate errors that leave students with more to debug than when they first encountered an error \citep{ConditionsLearningNovices, ComparisonDebuggingBehaviourNoviceExpert, MAADSMethod, ThinkAloudNoviceDebugging, DebuggingNoviceSkilledProgrammers}. From the students' perspective, this does not just make a working program a difficult goal; being stuck with an erroneous program can evoke emotional distress that negatively impacts students' experiences of learning to program \citep{ProgrammingAssignmentsEmotionalToll, EDAProgrammingEmotions, AnxietyLearningToProgram}. Additionally, there is a lack of studies that have investigated school students' debugging behaviour in a TBPL, despite the increasing number of students learning to program in this setting (e.g., \citep{UKSecondaryComputingCurriculum, ScotlandTechnologiesCurriculum, Swedish2024Curriculum, BritishColumbiaADSTCurriculum, ProgrammingSkillsInLatinAmerica}). This is a gap which we fill in this study.

\section{Method}
To investigate school students' debugging behaviour in a TBPL, several classes of lower secondary students individually completed five Python debugging exercises in one of their computing lessons. The exercises were hosted on an online code editor that logged program snapshots each time students attempted to run them. These snapshots were then qualitatively analysed to understand the changes students made when debugging the programs. Students also completed a survey about their attitudes and emotions towards debugging, which we report on in a separate publication \citep{AnonymisedPreviousStudy}. This section reports the components of the study in more detail. The debugging exercises, survey, and study website are available to view in the study repository \citep{AnonymisedStudyRepository}.

\subsection{Participants}
Students were purposively sampled due to the study's focus on secondary students learning a TBPL. We focused on lower secondary education (grades 6-8, ages 11-14) as this is when students begin learning a TBPL in (anonymised country) \citep{AnonymisedUKSecondaryComputingCurriculum}. Sampling began by inviting local secondary school computing teachers to participate in the study and then verifying that they taught a TBPL to their lower secondary students. Three teachers from two schools accepted, both of which were located in a suburban area with relatively high socioeconomic indicators and above-average general achievement. The teachers were briefed on the study procedure and invited to provide feedback on it.

A total of 73 students from three classes participated, not including two students who did not complete the study. Students self-reported their gender and year group in the post-survey. 36 identified as male, 24 as female, three identified as other and 10 preferred not to disclose. All students were in grades 7-8 and therefore almost certainly aged 12-14. However, since students did not explicitly report their age, some may have been slightly outside this range. Consent from the student, their teacher, and a parent or guardian was given before participation in the study, in accordance with the procedure approved by our department's ethics committee (anonymised IRB review ID).

At the time of the study, students had been learning Python in school for a maximum of one year. They had also been taught a block-based programming language in their secondary schools for at least one year before this, as well as for several years during primary school. Regrettably, we did not collect data on students' programming experience outside of the classroom or any prior debugging instruction they had received, such as debugging strategies or error types. We acknowledge this is a limitation of our study procedure that reduces our ability to contextualise our findings.

\subsection{Procedure and Data Collection}
The study was run by the participating teachers in one of their hour-long computing lessons, without the authors present.

\subsubsection{Debugging Exercises}
Students' first task was to complete a set of five debugging exercises, listed in Appendix \ref{appendix-b}. Each exercise consisted of a short erroneous Python program, a description of the program's intended function, and the number of errors present in the program. We advised teachers to spend half an hour on the exercises, though this was ultimately up to them. Python was the chosen programming language due to its popularity in (anonymised country) secondary schools \citep{AnonymisedUKITSSurvey, AnonymisedProgrammingSecondaryEducationEngland} and because all participating teachers taught it.

\begin{figure*}[ht]
\includegraphics[width=0.975\textwidth]{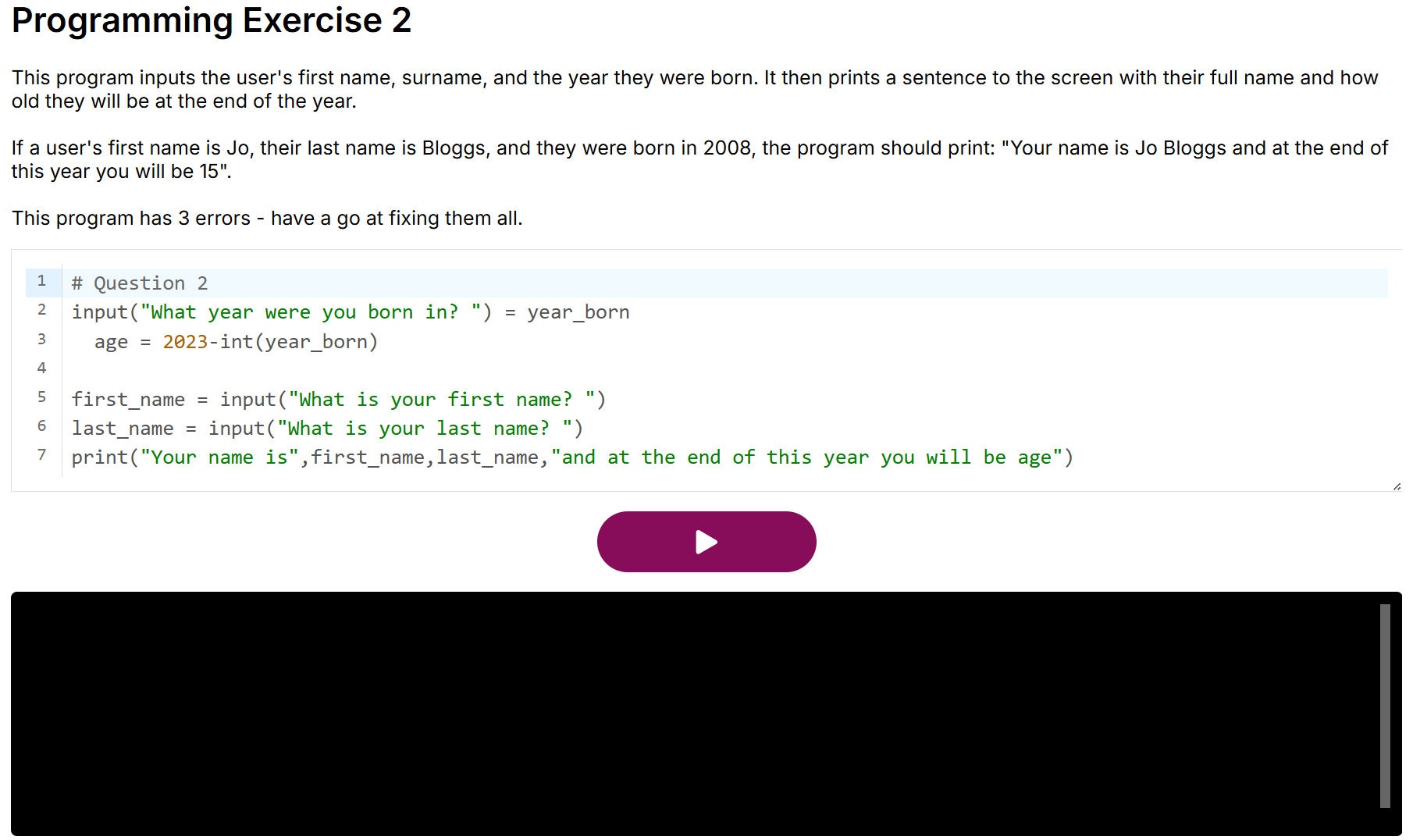}
    \caption{A programming exercise used in the study}
    \label{fig:debugging-exercise}
    \Description{A screenshot of a sample programming exercise, containing a program description, an erroneous Python program, and a lightweight code editor.}
\end{figure*}

The Python programs included a range of syntax, runtime, and logical errors that prior work has found common among beginner programmers \citep{PeaConceptualBugs, ErrorLandscapeCharacterising, ErrorMessageCausePython, MisconceptionsBeginnerProgrammer}. For clarity, we list definitions and examples of each error type in Table \ref{tab:error-type-definitions}. The exercises were developed by the authors, with the second author being an experienced computing educator. To ensure content validity of the debugging exercises, two participating teachers provided feedback on the exercises' appropriateness for their students, which was acted upon accordingly. The teachers were satisfied with the difficulty of the final exercises. Figure \ref{fig:debugging-exercise} displays one of the exercises. A breakdown of the number of errors originally present in each exercise is as follows:

\begin{itemize}
    \item \textbf{Exercise 1}: 1 syntax error, 1 runtime error.
    \item \textbf{Exercise 2}: 2 syntax errors, 1 logical error.
    \item \textbf{Exercise 3}: 1 syntax error, 3 logical errors.
    \item \textbf{Exercise 4}: 1 runtime error, 1 type error, 2 logical errors.
    \item \textbf{Exercise 5}: 1 syntax error, 3 logical errors.
\end{itemize}

\begin{table}[h]
    \caption{Definitions of error types}
    \label{tab:error-type-definitions}
    \begin{tabular}{p{1.4cm} p{6.65cm} p{5.85cm}}
        \toprule
        \textbf{Error type} & \textbf{Definition} & \textbf{Examples from the debugging exercises} \\
        \midrule
        Syntax & Errors that violate the grammatical rules of a programming language and prevent a program from executing. & Exercise 5 \begin{lstlisting}[style=tablelisting]
elif:
\end{lstlisting} \\
        Runtime & Errors that are raised during a program's execution due to a violation of the valid runtime behaviour of a programming language. & Exercise 1 (referring to \texttt{depth} before it is assigned a value) \begin{lstlisting}[style=tablelisting]
height = depth
depth = 25
\end{lstlisting}\\
        Type & Errors that violate the typing system of a programming language. These errors are raised during a program's execution for dynamically typed languages such as Python. & Exercise 4 \begin{lstlisting}[style=tablelisting]
count = "1"
while count<12:
\end{lstlisting} \\
        Logical & Errors that cause part of the actual program behaviour to misalign with its intended behaviour, as defined by a problem specification, but do not prevent the program from running its entirety. & Exercise 2 (the program is meant to print the \texttt{age} variable rather than the string \texttt{"age"}) \begin{lstlisting}[style=tablelisting]
print("Your name is",first_name,last_name,"and at the end of this year you will be age")
\end{lstlisting}\\
        \bottomrule
    \end{tabular}
\end{table}

Note that we explicitly distinguish between runtime and type errors. Being dynamically typed, type errors are an instance of a runtime error in Python. However, we perceive them as semantically distinct and hence categorise them separately; students who struggle to debug type errors may have different misunderstandings from those who struggle to debug other runtime errors.

We used an online, custom-built code editor that students had not used before to easily control the data we collected. The editor was minimal in its design to facilitate ease of use, containing only a text editor, a run button, and an output terminal (see Figure \ref{fig:debugging-exercise}). The code editor ran the programs using Skulpt\footnote{\url{https://skulpt.org/}}, an in-browser implementation of Python. At the time of the study, Skulpt raised different error messages than a typical Python interpreter, some of which were notably vague. We discuss this limitation in Section \ref{sec:varying-success-localisation}.

To enable the analysis we performed, the code editor logged the following data each time a student attempted to run a program:

\begin{itemize}
    \item The exercise the student was attempting.
    \item A snapshot of the program.
    \item Whether the code successfully executed (ran in its entirety without throwing an error message). If unsuccessful, the error message thrown by the program was also stored.
    \item A timestamp of when the program was run.
\end{itemize}

This data was saved to a database when a student had completed an exercise attempt. While the code editor also logged keystroke data, this was too fine-grained to sufficiently and reliably analyse for the number of exercises students had attempted. Although this meant we could not see the changes students made and reverted between runs, students would have had to run the program if they wanted to test their changes.

Students were not given feedback on the number of errors they resolved, nor were there any test harnesses for the programs. Instead, as an important part of the debugging process, students were encouraged to test the code frequently and move on to the next task when they wished. As we were interested in students' individual debugging behaviours, students were instructed not to collaborate or ask for their teacher's help, though they may have done so in practice.

\subsubsection{Attitudes and Emotions Survey}
Immediately after attempting the debugging exercises, students completed a short survey on their attitudes and emotions towards debugging\footnote{The full survey is available to view in the study repository \citep{AnonymisedStudyRepository}.}. The survey began with the following two questions on students' perceived performance and the debugging techniques they used across the exercises. 

\begin{itemize}
    \item \textit{``How well do you feel you performed when solving the errors in the programming exercises?''}
    \item \textit{``What techniques did you use to find and fix the errors in the programming exercises?''}
\end{itemize}

\begin{figure}
    \centering
    \includegraphics[width=0.9\linewidth]{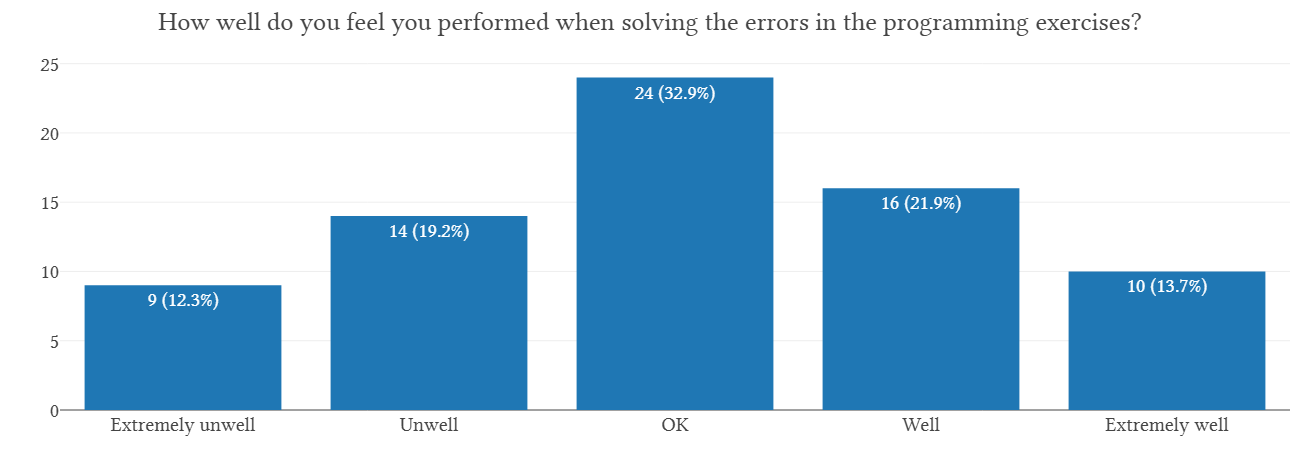}
    \caption{Students' perceived performance on the debugging exercises}
    \label{fig:perceived-performance-responses}
\end{figure}

Students answered the perceived performance question on a five-point Likert scale from `extremely unwell' to `extremely well', and their debugging techniques in a free-text response. Figure \ref{fig:perceived-performance-responses} visualises students' perceived performance. These were followed by 12 statements related to several attitudes and emotions towards debugging, which students responded to on a five-point Likert scale ranging `strongly disagree' to `strongly agree'. Students generally reported neutral or positive attitudes and emotions towards debugging, with several of them strongly correlated with self-efficacy \citep{AnonymisedPreviousStudy}. There were also statistically significant gender differences in debugging self-efficacy. The survey ended with demographic questions of students' school, gender, and year group.

\subsection{Analysis}\label{sec:analysis}
We performed qualitative content analysis (QualCA) \citep{QualContentAnalysisMayring} to analyse the changes that students made in the debugging exercises. This is a specific form of content analysis, a body of methods for extracting meaning from forms of communication in a rigorous and repeatable manner \citep{ContentAnalysisGuidebook, ContentAnalysisIntroMethodology, QualContentAnalysisMayring}.

Content analysis (CA) was a suitable body of methods for investigating our research question for two main reasons. First, CA is concerned with summarising data rather than interpreting it \citep{ContentAnalysisGuidebook}. This suits the snapshot data collected, which permits analysis of \textit{how} students changed an erroneous program across a set of runs, but not \textit{why} they made these changes. Where the former relates to students' behaviours, the latter relates to their strategies \citep{InteractionDomainStrategicKnowledge}, which cannot be inferred from program snapshots alone. While students' survey responses included limited data on their self-reported debugging `techniques', they were generally too short and vague to provide any meaningful insight.

Second, CA is applicable to all forms of data that contain meaning \citep{ContentAnalysisGuidebook, ContentAnalysisIntroMethodology, QualContentAnalysisMayring}. While individual snapshots provide little scope for analysis, a sequence of snapshots is a meaningful body of text that expresses a set of changes students make when debugging. Examining these can be used to analyse debugging behaviour. QualCA specifically was used due its focus on inductive category formation \citep{QualContentAnalysisMayring}, which was necessary due to a lack of research into secondary students' debugging behaviour in a TBPL.

\begin{figure}
    \centering
    \includegraphics[width=0.95\linewidth]{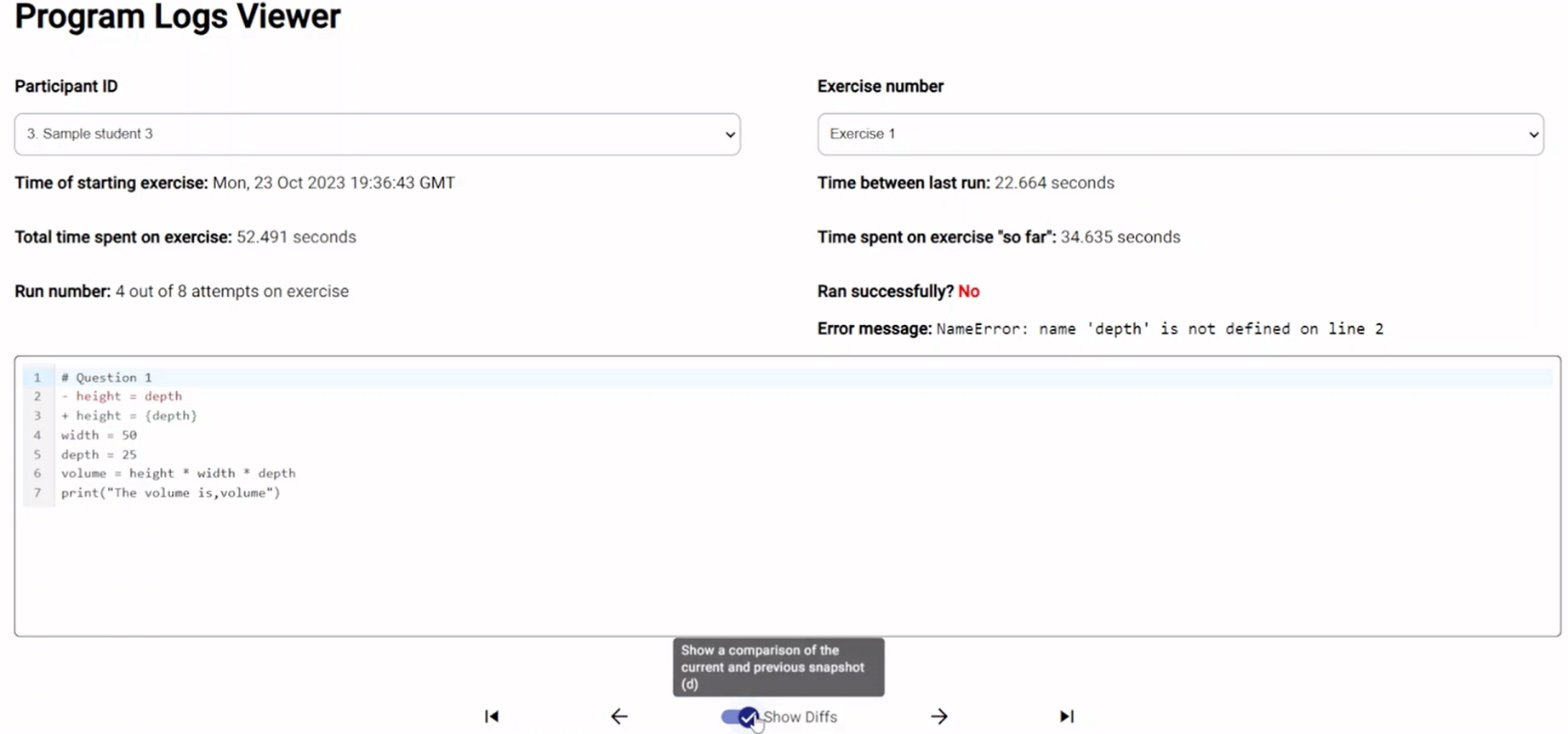}
    \caption{The analysis tool we used to analyse students' debugging behaviour, with coding and memoing functionality omitted}
    \label{fig:snapshot-replayer}
\end{figure}

QualCA was conducted through extensive manual inspection of students' exercise attempts, which primarily focused on the changes students made between runs. This involved an open-source tool\footnote{\url{https://anonymous.4open.science/r/program-logs-viewer-8F40}} that replayed students' snapshots for each exercise run-by-run, developed by the first author and displayed in Figure \ref{fig:snapshot-replayer}. The tool also displayed other contextual information, such as the total time on an exercise attempt, the time between runs, and the error message if one was raised for a given snapshot. Coding and memoing functionality was also included through checkboxes and input boxes (omitted from Figure \ref{fig:snapshot-replayer}). This approach enabled a detailed observation of students' debugging behaviours that was not possible with pure quantitative analysis, which other researchers have valued \citep{JadudCompilationBehaviour, WritingBetweenLines}.

The analysis began with the first author using this tool to view the exercise attempts for an extended period of time. This served two purposes: first, to consider what claims the analysis could soundly make and second, to inductively generate descriptive open codes. These and later higher-level codes were assigned to a specific exercise attempt rather than to a particular student. The first author then abstracted the open codes into more generalisable axial codes through further manual examination of the data. This led to low-level codes within the categorisation that appear quite objective, but ultimately summarise students' granular debugging behaviours. Higher-level categories were progressively refined through discussions between the authors and re-examination of the log data. This culminated in the final categorisation reported in Section \ref{sec:results} and Appendix \ref{appendix-a}.

Throughout the QualCA, each exercise attempt was coded with one or more of the lowest-level codes in the categorisation, as a student could display many different debugging behaviours. The higher-level categories were used only to organise the categorisation. The final categorisation consequently contained counts of each low-level code, which were aggregated for each higher-level category. To support our findings from the content analysis, we also conducted quantitative analysis of the dataset that is reported throughout Section \ref{sec:results}.

The second author used an intermediate version of the categorisation to code small samples of exercise attempts at two points. The first of these was at the end of the open coding stage, where the second author coded five exercise attempts to gauge the scope of the coding system. They then coded 10 attempts after the initial categorisation was formed. The categorisation was refined after both of these.

\subsection{Difficulties in Consistently Coding with the Categorisation}\label{sec:difficulty-agreeing-categorisation}
At this point, we had a `final' version of the categorisation which we wanted to assess the intercoder reliability (ICR) of. This was necessary as we wanted to compare the frequencies of debugging behaviours \citep{QualContentAnalysisMayring, IRRHCI}. In practice, however, we found it difficult to achieve sufficient levels of agreement for some categories. The following sections detail our ICR procedure and how the low levels of agreement affect the reliability of our results. We reflect on qualitatively analysing program snapshot data more generally in Section \ref{sec:limitations}.

\subsubsection{Intercoder Reliability Procedure}
Two ICR rounds took place, the first of which was between the authors on a random sample of 45 exercise attempts (14\% of the total attempts). At this point, the second author had used earlier versions of the categorisation to code small samples of exercise attempts but had not seen the final version of the categorisation developed by the first author.

Before coding, the first author introduced the second author to the definitions for the codes in the categorisation and modelled an example of coding an exercise attempt. After the authors had both independently coded the sample of exercise attempts, we originally calculated the ICR of the categorisation using Cohen's kappa ($\kappa$) \citep{CohensKappa}. This yielded a value of $\kappa = 0.64$, indicating moderate agreement \citep{IntervalEstimationCohensKappa}. Based on this, the first author then coded the rest of the exercise attempts using the categorisation.

However, after completing the coding, we realised that $\kappa$ was not a valid metric for assessing the ICR of the categorisation. The exercise attempts were coded with multiple codes per category, making our data \textit{multi-valued}, and $\kappa$ can only be used for single-valued data\footnote{Even then, \citeauthor{ReliabilityQualAnalysisMisconceptions} describes $\kappa$ as `just about
worthless as a reliability index in content analysis' \citep[p. 422]{ReliabilityQualAnalysisMisconceptions} as it can increase even when disagreement between two coders \textit{increases}.} \citep{IntercoderReliabilityMisuse, CohensKappa}. As a result, we recalculated the ICR of the categorisation using Krippendorff's multi-value alpha ($_{mv}\alpha$) \citep{KrippendorffMultiValueAlpha}, a variant of Krippendorff's alpha \citep{ReliabilityContentAnalysisBook} for multi-valued data, using the \texttt{mvalpha} package in R \citep{mvAlphaPackage}. We calculated $_{mv}\alpha$ per-category to give a more accurate breakdown of the categorisation's ICR than a single value, as recommended by \citet{ReliabilityContentAnalysisBook} and \citet{IntercoderReliabilityMisuse}.

\begin{table}[h]
    \caption{The $_{mv}\alpha$ values for each intercoder relaibility round}
    \label{tab:mv-alpha-values}
    \centering
    \begin{tabular}{c c c}
        \toprule
        \textbf{Category} & \textbf{Round 1} & \textbf{Round 2}  \\
        \midrule
        First move & 0.90 & 0.82 \\
        Positive debugging behaviours & 0.41 & 0.41 \\
        Added errors & 0.42 & 0.49 \\
        Resolved errors & 0.50 & 0.19 \\
        Inconsequential changes & 0.52 & 0.18 \\
        Miscellaneous changes & 0.58 & 0.43 \\
        \bottomrule
    \end{tabular}
\end{table}

This recalculation yielded $_{mv}\alpha$ values between 0.41--0.90 (see Table \ref{tab:mv-alpha-values}). \citet{KrippendorffMultiValueAlpha} do not state guidelines for acceptable values for $_{mv}\alpha$, but \citet{ReliabilityContentAnalysisBook} suggests elsewhere a $\alpha$ (single-valued) $ > 0.67$ for categories used in exploratory research. Although this would have been harder to achieve with $_{mv}\alpha$ and our categorisation, where complete agreement required an exact match of several codes between coders, the low $_{mv}\alpha$ values prompted us to perform another ICR round.

The second ICR was conducted between the first author and a research colleague on a different random sample of 24 exercise attempts (7.5\% of the total attempts). We did not refine the categorisation as this version had been used to code all of the exercise attempts. Before independently coding the exercise attempts, the first author showed their colleague the categorisation, the code definitions, a worked example of coding an exercise attempt, and a coding flowchart to support replicable coding. However, this ICR round generally yielded lower $_{mv}\alpha$ values (0.19--0.82) than the first (see Table \ref{tab:mv-alpha-values}).

\subsubsection{Implications for Results}
Our ICR procedure show that human coders could not consistently apply most of the categorisation, which limits the reliability of our results. We believe this is mostly due to the structure of the categorisation; each category contained 2--25 low-level codes, many of which could be applied to a single exercise attempt. These codes were deliberately low-level to remain descriptive rather than interpretative, but resulted in a large categorisation that was difficult to apply without extensive familiarity and immersion in the data. Unsurprisingly, the more codes a category contained, the lower level of agreements we generally obtained. We also believe the lack of agreement is partly due to the general difficulty of coding program snapshot data, which we discuss further in Section \ref{sec:limitations}.

Nonetheless, we still believe our content analysis provides novel insights into students' debugging behaviour in Python. We therefore provide a qualitative account of the categorisation in the following section, supplemented with anonymised student snapshots and quantitative statistics. We do not report the frequencies of each code in the next section due to the low levels of agreement\footnote{Apart from the \textit{first move} category, due to the substantial agreement we achieved.}, but do include them in Appendix \ref{appendix-a}  for transparency. The source code used for calculating the $_{mv}\alpha$ vallues is also available in the study repository \citep{AnonymisedStudyRepository}.

\section{Results}\label{sec:results}
A variety of debugging behaviours were present across the 322 exercise attempts, represented under the following six categories below. Readers interested in the full categorisation should refer to Appendix \ref{appendix-a}.

\begin{enumerate}
    \item \textbf{First move:} Whether a student ran their program before making changes to it, or vice versa.
    \item \textbf{Positive debugging behaviours:} Behaviours that facilitated successful debugging.
    \item \textbf{Added errors:} Changes that introduced errors not original present in the programs.
    \item \textbf{Resolved errors:} Changes that removed an error(s) originally present in the programs.
    \item \textbf{Inconsequential changes:} Changes that had no impact on the state of the programs.
    \item \textbf{Miscellaneous behaviours:} Other changes that did not fit into the above categories.
\end{enumerate}

Table \ref{tab:log-data-summary-stats} provides some higher-level statistics that support the behaviours observed through manual inspection; most students did not successfully resolve all of the errors in the debugging exercise and often made quick and small-scale changes between runs. The median Levenshtein distance\footnote{The number of removals, additions, or substitutions required to transition from one string to another.} between consecutive runs was 2, not including the 4,809 identical consecutive runs.

\begin{table}[h]
 \caption{Summary statistics for each debugging exercise}
 \label{tab:log-data-summary-stats}
    \begin{tabular}{c p{1.2cm} p{1.2cm} p{2.7cm} p{2.7cm} p{4cm}}
        \toprule
         \textbf{\#} & \textbf{$n_{attempts}$} & \textbf{$n_{correct}$} & \textbf{Median $n_{runs}$} & \textbf{Median time ($s$)} & \textbf{Median Levenshtein distance between non-identical consecutive snapshots} \\
         \midrule
            1 & 72 & 34 & 7 ($skew = 2.54$) & 153 ($skew = 2.33$) & 2 ($skew = 2.07$) \\
            2 & 71 & 36 & 13 ($skew = 1.38$) & 304 ($skew = 1.30$) & 3 ($skew = 4.76$) \\
            3 & 68 & 25 & 13 ($skew = 1.48$) & 343 ($skew = 1.54$) & 2 ($skew = 5.84$) \\
            4 & 60 & 23 & 13 ($skew = 1.35$) & 251 ($skew = 1.15$) & 2 ($skew = 29.98$) \\
            5 & 51 & 24 & 35 ($skew = 1.56$) & 290 ($skew = 1.10$) & 3 ($skew = 4.2$) \\
        \midrule
        \textbf{Overall} & 322 & 142 & 13 ($skew = 3.33$) & 274 ($skew = 1.61$) & 2 ($skew = 7.99$) \\
         \bottomrule
    \end{tabular}
\end{table}

Table \ref{tab:end-state-breakdown} shows the end state of the debugging exercises, derived through manual inspection of each attempt's final snapshot. A correct program was defined as one that included no errors and outputted what was specified in the exercise descriptions. Out of the attempted exercises, 44\% ended in a correct state. Eight students (11\%) ended every attempted exercise in a correct state, compared to 26 students (36\%) who ended every attempted exercise with at least one error. We address the difficulty of the exercises in Section \ref{sec:fragile-knowledge}.

\begin{table}
 \caption{The final state of each debugging exercise attempt, categorised through manual inspection. A snapshot may either contain one or more error types or be in a correct state. The underlined cells represent error types that were originally present in the exercises. Percentages are expressed in terms of the number of attempts at the exercise.}
 \label{tab:end-state-breakdown}
 \centering
    \begin{tabular}{c c c c c c c c}
        \toprule
         \textbf{\#} & \textbf{Syntax errors} & \textbf{Runtime errors} & \textbf{Type errors} & \textbf{Logical errors} & \textbf{Correct state} & \textbf{Total attempts} \\
         \midrule
            1 & \textul{7 (10\%)} & \textul{14 (19\%)} & 2 (3\%) & 27 (38\%) & 34 (47\%) & 72 \\
            2 & \textul{27 (38\%)} & 5 (7\%) & 4 (6\%) & \textul{25 (35\%)} & 36 (51\%) & 71 \\
            3 & \textul{21 (31\%)} & 0 & 2 (3\%) & \textul{42 (62\%)} & 25 (37\%) & 68 \\
            4 & 12 (20\%) & \textul{16 (27\%)} & \textul{23 (38\%)} & \textul{35 (58\%)} & 23 (38\%) & 60 \\
            5 & \textul{8 (16\%)} & 1 (2\%) & 0 & \textul{26 (51\%)} & 24 (47\%) & 51 \\
            \midrule
            \textbf{Overall} & 75 (23\%) & 36 (11\%) & 31 (10\%) & 155 (48\%) & 142 (44\%) & 322 \\
         \bottomrule
    \end{tabular}
\end{table}

We now report on the categorisation in depth, referencing sub-categories in italics and pseudonymous students' snapshots in listings\footnote{Green and red highlighted sections indicate additions and removals compared to the original state of the program.} where appropriate. We use the term \textit{successful execution} refers to programs that run in their entirety without throwing an error. This does not include programs that throw an error message during their execution.

\subsection{First Move}\label{sec:first-move}
Students demonstrated two behaviours when they began the exercises. They either \textit{ran the program before making changes} (ran-first), or \textit{made changes before running the program} (changed-first).

Most exercise attempts (66\%) began by running first, which often happened soon after beginning the exercise. The median run-first time was 17 seconds ($skew = 4.26$); likely not enough time to read the exercise description or comprehend the program. Although students may have quickly ran-first to view any error messages, the median time between their first and second runs was 22 seconds ($skew = 4.38$).

Naturally, students who changed-first tended to first run their program later on (median = 63 seconds, $skew = 1.47$). These students did not have any error messages to guide their initial edits. 84\% of these attempts still failed to run the first time. 

\subsection{Positive Debugging Behaviours}\label{sec:positive-debugging-behaviours}
After their first run, 94\% of programs did not successfully execute and displayed an error message. Many of the changes that immediately followed were made on the line specified by the error message (\textit{first change on line referred to in error message}). Although this behaviour generally helped students to narrow down the search space, error messages were sometimes misleading. For instance, exercise 4 originally had a type error on line 2, but the error message referred to line 4 (see Appendix \ref{sec:appendix-exercise-4}). As a result, 31 students (52\% of attempts on the exercise) made non-corrective changes to line 4 after running-first. Furthermore, if students did make changes to the erroneous line, it did not guarantee that they truly localised an error. As the following categories highlight, students made haphazard changes to parts of the erroneous lines that were originally correct.

A small number of students also \textit{made improvements to the program}, such as by adding exceptions to handle particular inputs, or \textit{entered incorrect input}, which was reflected in the error messages that were thrown.

\subsection{Added Errors}\label{sec:introduced-additional-errors}
Many students introduced errors that were not previously present in the programs. In particular, students \textit{added syntax errors} through a variety of behaviours: output statements, variable assignments, and symbols were all incorrectly syntactically edited through the addition or removal of a few characters.

Leandro's attempt to exercise 3 is a good example. Leandro rated his performance on the exercises as ok and explicitly reported using \textit{``trial and error''} in the survey. After running-first 22 seconds into the exercise, Leandro added a colon after the \texttt{if} keyword on the line referred to in the error message (\textit{incorrect syntactical changes involving selection}), highlighted in the listing below. This adds a syntax error to the program; even if Leandro resolved the original error on line 6 by changing the assignment operator (=) to an equality operator (==), he would still receive the same error message when running the program.

\begin{lstlisting}[style=studentsnapshotlisting, title=Leandro's second run on exercise 3. Error message: \texttt{SyntaxError: bad input on line 6}., escapechar=!]
# Question 3
print("This program will check if you should apply to be a computing teacher")
age = int(input("What is your age? "))
computing_degree = input("Do you have a passion for teaching computing? Enter 'yes' or 'no': ")

if!\colorbox{green}{:}! age > 21 or computing_degree = "yes":
  allowed_to_apply = "Successful"
else:
  allowed_to_apply = "Unsuccessful"
  print("Result of check:",allowed_to_apply)
\end{lstlisting}

Some changes that added errors could be considered `slips' as they were quickly fixed in the following runs. However, students were in a syntax error state for a median of 88 seconds per exercise attempt ($skew = 1.59$), suggesting that some syntax error additions were not simple to resolve. As Leandro's change demonstrates, worsening the syntactic correctness of an already erroneous line was often not reflected in the error message.

A common behaviour that added a runtime error was \textit{changes to variable assignments}, which was often made by converting an assignment operator (\texttt{=}) to an equality operator (\texttt{==}). Type errors added also frequently involved variable assignments or references, typically by erroneously removing or adding casts.

Students also made illogical edits to their programs, often when they were not successfully executing. Common incorrect changes involved \textit{variable assignments}, \textit{logical or mathematical operators} and \textit{program flow}, often by manipulating the program's indentation.

Beth, who reported performing extremely well and did not meaningfully respond to the question on debugging techniques, made an erroneous change to the program's flow in her attempt to exercise 5. She completely removed the syntactically invalid line 16 in her eighth run, allowing the program to execute successfully. The consequence of this change, however, was logically incorrect. Even if every other logical error were resolved, there would be no case where \texttt{"The winner is Player 2"} is correctly printed. This incorrect change persisted for the remainder of her 32 runs.

\begin{lstlisting}[style=studentsnapshotlisting, title=Beth's eighth run on exercise 5. The program successfully executed., escapechar=!]
# Question 5
import random
player_1 = random.choice(["Rock","Paper","Scissors"])
print("Player 1 has chosen "+player_1)
player_2 = random.choice(["Rock","Paper","Scissors"])
print("Player 2 has chosen "+player_2)

if player_1 == player_2:
  winner = "Neither"
elif player_1 == "Rock" and player_2 == "Scissors":
  winner = "Player 1"
elif player_1 =="Paper" or player_2 == "Rock":
  winner = "Player 1"
elif player_1 == "paper" or player_2 == "Paper":
  winner = "Player 1"
!\colorbox{red}{elif:}!
  winner = "Player 2"
  print("The winner is neither")
\end{lstlisting}

\subsection{Resolved Errors}\label{sec:resolving-errors-categories}
Students also resolved errors through various behaviours. This category only accounts for errors originally present in the debugging exercises due to the complexity in manually detecting errors that were introduced and then resolved. The main type of resolution were changes that entirely removed the error in question, such that the erroneous part of the program was completely correct after the change, which we titled \textit{correct resolutions}.

Students also \textit{added} logical errors through their resolutions of syntax, runtime, and type errors, as demonstrated by Bukayo's attempt to exercise 1. Bukayo rated their performance as extremely unwell and responded to the debugging techniques question with \textit{``i ran code and it said unsuccessful''}. After running-first 30 seconds into the exercise, Bukayo inserted a quotation mark in the green highlighted position on line 6, where the error message referred to. In doing so, Bukayo \textit{introduced a logical error from the resolution of a syntax error}, as the program printed the string \texttt{"volume"} rather than the value of the \texttt{volume} variable. This particular first change on exercise 1 was exhibited by 14 other students (19\% of attempts on the exercise). While these changes could be intermediary edits that help students get closer to a correct program, they were not always removed.

\begin{lstlisting}[style=studentsnapshotlisting, title=Bukayo's second run on exercise 1. Error message: \texttt{NameError: name \textquotesingle depth\textquotesingle \  is not defined on line 2}, escapechar=!]
# Question 1
height = depth
width = 50
depth = 25
volume = height * width * depth
print("The volume is!\colorbox{cyan}{"}!,volume!\colorbox{green}{"}!)
\end{lstlisting}

The point at which a logical error was correctly resolved, relative to the program state, was also categorised through manual inspection. Interestingly, students often \textit{resolved a logical error while the program was not running}, while some resolved logical errors after one or more successful executions, usually in quick succession. This was particularly prevalent for exercises containing no input statements. We discuss repeated runs more in Section \ref{sec:misc-behaviour}.

Errors were also resolved in a \textit{hard-coded} fashion. These represent unrobust resolutions that satisfied the exercise descriptions but would be incorrect if the programs were tested with multiple variable values. Frida's attempt to exercise 1 contains an example of a hard-coded change. They rated their performance as extremely unwell and \textit{``ran the code and see the errors''} and hard-coded the \texttt{height} variable in their final run.

\begin{lstlisting}[style=studentsnapshotlisting, title=Frida's third and final run on exercise 1. The program successfully executes., escapechar=!]
# Question 1
height = !\colorbox{green}{25}!
width = 50
depth = 25
volume = height * width * depth
print(f"The volume is {volume}")
\end{lstlisting}

\subsection{Inconsequential Changes}
Some changes that students made had no effect on the state of their programs. That is, they did not resolve errors in the program or add any. Consistent with many of the other behaviours, these inconsequential changes were often no more than a few characters, and commonly related \textit{to the whitespace of the program}, \textit{to existing output statements}, or \textit{involved }[Python]\textit{ symbols}.

Kai's attempt to exercise four exhibits several inconsequential changes early on. Kai reported performing well on the exercises, and \textit{``ran the code then kinda thought about it as if what i would do if i did it from scratch''}. After running-first 63 seconds in, Kai made three inconsequential changes in his following three runs, highlighted in green below. Although each of Kai’s changes was to erroneous lines, none of them added or resolved any errors, despite taking almost two and a half minutes to make.

\begin{lstlisting}[style=studentsnapshotlisting, title=Kai's fourth run on exercise 4. Error message: \texttt{TypeError:\textquotesingle <\textquotesingle \ not supported between instances of \textquotesingle str\textquotesingle \ and \textquotesingle int\textquotesingle \ on line 4.}, escapechar=!]
# Question 4
count!\colorbox{green}{ }!=!\colorbox{green}{ }!"1"
print("The first 12 multiples of the number 6 are:")
while !\colorbox{green}{(}!count!\colorbox{green}{) }!<!\colorbox{green}{ }!12:
  times_table = 6 * count
  count=count+1
print(times_Table)
\end{lstlisting}

Some inconsequential changes may have aided students' program comprehension. For example, students may have benefitted from separating different parts of the program with whitespace or by adding comments. From our manual inspection, however, such changes were less common than seemingly unsystematic edits.

\subsection{Miscellaneous Behaviours}\label{sec:misc-behaviour}
A widespread behaviour was the repeated running of identical programs. This is easily measured automatically; if two or more consecutive snapshots are identical, the student has repeatedly run their program without making changes or reverted any changes they did make between runs. There were
4,809 snapshots that were identical to their previous snapshot, accounting for 67\% of all runs. Despite its prevalence, only 10 students mentioned repeatedly running code in the survey \citep{AnonymisedPreviousStudy}.

The slightly more common version of this behaviour was when students’ programs were successfully executing, accounting for 2,526 runs (35\% of all runs). Only 265 (12\%) of these identical runs were over five seconds apart, while 86\% of them occurred in exercise five, a program that contains no input statements. Such short times between runs would not have given students time to comprehend and verify the output's correctness.

Students also repeatedly ran their programs when they contained errors, accounting for 2,283 (32\%) runs. The time between these runs was low and heavily skewed (median = 0.367 seconds, $skew = 7.86$). When students were running identical programs in such quick succession, the error message would have remained the same, and they would not have had time to enter any inputs into the program.

Finally, students often \textit{reverted previous changes} they made earlier in the debugging exercises and also \textit{reverted corrective changes}, hence re-adding errors that they had previously resolved.

\section{Discussion}\label{sec:discussion}
The categorisation evidences a wide range of debugging behaviours exhibited by students. While most students were able to debug some errors, many students' changes did not resolve errors and sometimes moved them \textit{further away} from a correct program. We now summarise four common behaviours that we observed from our manual inspection and supplementary quantitative analysis, which we use to consider barriers preventing students from successfully and reliably debugging in a TBPL.

\subsection{Common Observations}

\subsubsection{Early Changes}
Most exercise attempts began with one or more runs soon after the start of the exercises, quickly followed by changes to the program (see Section \ref{sec:first-move}). Given the time between these early runs was often short, this behaviour suggests that students commonly attempted a fix before they had given themselves time to understand the exercise description, comprehend the program, or understand the issue with the program. In other words, these results imply a lack of engagement with the initial stages of the debugging process, which likely impedes students' ability to successfully locate and resolve errors \citep{AnalysingIntroductoryDebuggingProcess}. Previous studies of school students' debugging behaviour have found a similar tendency \citep{ConditionsLearningNovices, ProblemSolvingDebuggingK68, ProblemSolvingGroupsLearningProgramming}, which does not allow them to view the program as a functional whole. This is particularly problematic when students are debugging programs they are not familiar with \citep{BugLocationStrategiesAnalysis}, which is likely when working with peers or programming with generative AI. Even if this is not the case, programming lessons in schools are often spaced up days or weeks apart, meaning students will likely have to comprehend programs they have previously written. 

The prevalence of early changes may have been affected by the design of the debugging exercises, most of which contained simple syntax errors. Students may have resolved these errors early on, then comprehended the program in detail to resolve the logical errors. This effective approach to debugging was too high-level for our categorisation to capture. However, the common short times between initial runs and struggles with resolving syntax errors suggest that this was not always the case.

\subsubsection{Varying Success with Error Localisation}\label{sec:varying-success-localisation}
At least some of students' initial changes to the erroneous programs were on the line specified in the error message thrown at the beginning of the exercises (see Section \ref{sec:positive-debugging-behaviours}). This behaviour partially explains the early running-first discussed in the previous section; students may have wanted their changes to be guided by some targeted feedback. However, students' ability to make changes on erroneous lines was often coupled with an inability to resolve errors. This makes it difficult to tell whether students had precisely located errors, as many changes were made to non-erroneous parts of erroneous lines, or whether they lacked the necessary programming knowledge to make a correct fix. \citeauthor{NottinghamDebuggingBehaviour}'s study of CS1 students found that two thirds of `good programmers' were able to find but not fix errors \citep{NottinghamDebuggingBehaviour}, while \citet{ExploringCollabDebuggingProcesses} found that most school students `implicitly' localised errors in their physical computing projects. However, most studies of school students' debugging behaviour do not specifically discuss difficulties with localisation.

The ambiguity of the error messages raised by the code editor in our study likely affected this pattern. `\texttt{Bad input on line 2}' does not reveal any information other than the erroneous line, which itself was not always accurate. Although efforts to enhance error messages have reported conflicting effectiveness (e.g., \citep{EnhancingSyntaxErrorsAppearsIneffectual, EnhancedCEMsResultsInconclusive, NotSilverBulletLLMPEMs, EffectsEnhancedPEMsOnSyntaxDebuggingTest, EffectiveApproachErrorMessageEnhancement}), improving their specificity for the code editor we used would have helped us to more accurately identify whether students were struggling to locate or resolve errors.

\subsubsection{Tinkering and Extreme Moving}
Our manual inspection of students' debugging behaviours found that many students' changes were small-scale, quickly made, and sometimes reverted. This cycle of repeated rapid-fire changes is reminiscent of tinkering and extreme moving \citep{ConditionsLearningNovices}, which has frequently been reported in studies of school students' debugging behaviour \citep{ProblemSolvingDebuggingK68, ElementaryPuzzleBasedDebugging, NoviceDebuggingVBProgramming, ElementaryGameBasedDebugging}. While most of these studies have involved game-based programming environments, our findings show that similar behaviours exist when school students are debugging text-based programs. More specifically, our categorisation shows that many of these behaviours added syntax errors or made no difference to the correctness of the program.

If students use behaviours in line with tinkering and extreme moving when debugging their own programs, a successfully executing program becomes an increasingly difficult goal. Students are unlikely to consistently revert every erroneous change they make, which means that additional errors are likely to accumulate \citep{ConditionsLearningNovices, ComparisonDebuggingBehaviourNoviceExpert, MAADSMethod, ThinkAloudNoviceDebugging, DebuggingNoviceSkilledProgrammers} and corrective changes may be reverted \citep{ProblemSolvingDebuggingK68, ConditionsLearningNovices}. This sort of ``impulsive, unreflective coding'' can easily ``leave some movers moving without getting anyplace.'' \citep[p. 44]{ConditionsLearningNovices}. If multiple students are tinkering or extreme moving in a classroom environment, teachers may be inundated with students who are stuck with many errors in their programs \citep{CurrentPerspectivesDebugging, HandsUpProblem}, reducing the likelihood that students will receive the support they are in need of. Again, some of this behaviour may have been influenced by the debugging exercises, some of which only required a small change to resolve an error.

\subsubsection{Successful Resolutions}
From our manual inspection, we observed that when students exhibited behaviours in line with tinkering and extreme moving, they were generally unsuccessful at resolving errors. However, the vast majority of students did resolve some errors in the debugging exercises. Rather than employing effective behaviours, these students tended not to resort to the ineffective behaviours that prevented successful debugging. This is reflected in the relative lack of behaviours in the \textit{resolved errors} category, some of which also added errors to the program.

The way we analysed students' exercise attempts may have yielded a higher prevalence of ineffective behaviours. There are a limited number of ways to correct the syntax of ‘\texttt{if computing\_degree = "yes":}’; there are a large number of ways to incorrectly alter it. However, other studies have associated debugging success with smaller and more targeted changes \citep{MAADSMethod}, which would yield a smaller range of behaviours than when unreflectively debugging.

\subsection{Barriers to Successful and Reliable Debugging}\label{sec:barriers-effective-debugging}
The range of behaviours we observed begs the question: ``what barriers are preventing more students from successfully and reliably debugging?'' Both success and reliability are important; rapid-fire, unreflective, and speculative changes sometimes work, but are not robust approaches for consistently resolving errors. Instead, students should be able to apply effective debugging strategies in a range of situations, which is not always the case \citep{NottinghamDebuggingBehaviour}. This section lists four barriers to successful and reliable debugging and discusses how teachers and researchers can work to alleviate these.

\subsubsection{Fragile Knowledge}\label{sec:fragile-knowledge}
Students often struggled to fix errors, even when they were making changes to erroneous lines. When students had successfully located an error(s), the inability to resolve it may have been due to \textit{fragile knowledge} of the programming constructs they were attempting to fix. As mentioned in Section \ref{sec:related-work}, fragile knowledge is ``knowledge that is partial, hard to access, and often misused'' \citep[p. 4]{FragileKnowledgeOriginalPaper} and can be broken into the following four types:
\begin{itemize}
    \item \textbf{Partial knowledge:} Concepts that students do not fully understand, possibly because they have not yet been taught them.
    \item \textbf{Inert knowledge:} Knowledge that a student possesses but is unable to recall when required.
    \item \textbf{Misplaced knowledge:} Knowledge that a student recalls at the incorrect time.
    \item \textbf{Conglomerated knowledge:} Combining several programming elements in a semantically or syntactically incorrect manner ``in an attempt to provide the computer with the information it needs'' \citep[p. 7]{FragileKnowledgeOriginalPaper}.
\end{itemize}

Many of the ineffective behaviours we observed suggested some level of fragile knowledge. For example, many students added Python symbols to syntactically or logically invalid locations, suggesting misplaced knowledge, and incorrectly combined several constructs, suggesting conglomerated knowledge. More generally, frequent small-scale changes could imply any type of fragile knowledge. If students had sufficient knowledge of how to correctly implement an erroneous programming construct, they would not have to make many changes to fix it. If their knowledge is fragile, however, tinkering and extreme moving are methods of quickly experimenting with many potential fixes. Regardless of what type, \citeauthor{FragileKnowledgeOriginalPaper} state that ``fragile knowledge will not sustain any sophisticated problem-solving'' \citep[p. 28]{FragileKnowledgeOriginalPaper}.

The presence of fragile knowledge, and the general lack of success with the exercises, may imply that the debugging exercises were too difficult for students. We believe this is not the case for three reasons. First, the participating teachers were satisfied with the difficulty of them. Second, students' perceived performance on the exercise appeared normally distributed (see Figure \ref{fig:perceived-performance-responses}). Third, almost all students were able to resolve some of the errors in the challenges. Rather, as discussed in Section \ref{sec:related-work}, fragile knowledge may be a barrier to \textit{consistently} being able to resolve errors \citep{NottinghamDebuggingBehaviour, ThinkAloudNoviceDebugging}.

Fragile knowledge is an example of a general difficulty of learning to program which impacts debugging behaviour. If fragile knowledge is a barrier to error resolution specifically, teachers could provide scaffolded support for fixing errors in their interactions with students, such as hints or providing students with a fix \citep{HandsUpProblem, TeacherSupportForDebuggingPhysicalComputing, TeacherDiagnosticInterventionSkills}. Since they will not always have time for such interactions in the classroom \citep{CurrentPerspectivesDebugging}, pedagogical tools for debugging could provide similar tailored assistance to students struggling to fix errors. 

\subsubsection{Lack of Systematicity and Reflection}
Students often made changes so early on that it would not have been possible to properly comprehend the erroneous programs, understand their intended behaviour, or hypothesise about the cause of the errors. The frequently observed behaviours in line with tinkering and extreme moving are similarly characterised by a lack of reflection \citep{ConditionsLearningNovices}. This lack of systematicity likely prevented students from making correct fixes; reflecting on each stage of the debugging process would have helped them to make more intentional changes with a greater understanding of the program. Since we did not collect data on students' awareness of the debugging process or debugging strategies, possible drivers for this barrier could be a lack of knowledge of the debugging process, a lack of explicit instruction in effective debugging strategies, or a preference against using them.

Many pedagogical approaches for modelling a systematic debugging process have been developed by computing education researchers (e.g., \citep{CarverImprovingChildrensDebugging, CarverImprovingChildrensDebugging, AnalysingIntroductoryDebuggingProcess, ExplicitlyTeachingDebuggingPrimarySchool}), with some success. A recent meta-analysis of debugging interventions found systematic processes to have the second largest effect on debugging skills \citep{DebuggingMetaAnalysis}, but evidence that students adopt these approaches to their own debugging practice over time is lacking \citep{DebuggingInterventionLitReview}. Even when taught a systematic process, students may prefer to use unsystematic strategies \citep{TeachingExplicitProgrammingStrategies, NoviceReflectionsDebugging}.

To better understand the effect of systematic processes on students' debugging behaviour, longer-term intervention studies of systematic debugging are required \citep{DebuggingInterventionLitReview}. Investigating \textit{how} these approaches are delivered is also important. Purely instructional systematic processes (e.g., \citep{CarverImprovingChildrensDebugging, SystematicProcessMichaeli, AnalysingIntroductoryDebuggingProcess, ExplicitlyTeachingDebuggingPrimarySchool}) do provide a reliable model for students, but no way of enacting them \textit{within} a programming environment. While pedagogical tools for systematic debugging \citep{Ladebug, SystematicDebuggingLogicalErrors} contain useful features for encouraging systematicity, these are still separate from the programming environments that students will typically be debugging in. A more effective approach could be integrating features that encourage systematic debugging \textit{into} beginner programming environments. For example, students could be encouraged to write what they are going to change in the program before being able to attempt a fix \citep{ReflectiveDebuggingSpinoza}. Such features may more effectively help students to adopt a more systematic approach to debugging.

\subsubsection{The Syntax Barrier}
Much of the variation in students' debugging behaviour involved incorrect syntax. Many programs remained syntactically invalid for extended periods of time, and students commonly added syntax errors when attempting to debug other errors. These findings support the idea of the `syntax barrier' as a significant struggle for beginner programmers \citep{UnderstandingSyntaxBarrier}. This study specifically highlights secondary school students' struggles with resolving syntax errors in Python and the frequency with which they add them. 

There are several tools and approaches that can alleviate the syntax barrier for students. Programming error messages have the potential to be a useful source of detailed feedback for localising errors and suggesting fixes. Currently, however, most computing education research into enhanced PEMs has shown conflicting effectiveness and mainly involved undergraduate students (e.g., \citep{EnhancingSyntaxErrorsAppearsIneffectual, EnhancedCEMsResultsInconclusive, NotSilverBulletLLMPEMs, EffectsEnhancedPEMsOnSyntaxDebuggingTest, EffectiveApproachErrorMessageEnhancement}). Another way of reducing the syntax barrier is to use tooling that makes syntax errors near-impossible. Block-based languages do this, but at the cost of expressivity and potentially perceived utility \citep{ToBlockOrNotToBlock}. Alternatively, \textit{frame-based editing} is a paradigm for writing programs that combines elements of block- and text-based programming \citep{FrameBasedEditing}. Many syntax errors are simply not possible in FBEs, saving time that beginners spend dealing with these difficulties \citep{CostSyntaxHowToAvoidIt}. Using frame-based environments (e.g., \citep{StrypePoster2024, TeachingCodingDifferentlyElan}) in the classroom, particularly at the very beginning of students' journey with learning a TBPL, would allow students to successfully execute their program more often and consequently focus on the logic of their programs.

\subsubsection{Dynamics of Attitudes and Emotions}
Students' attitudes and emotions towards debugging in the post-survey varied \citep{AnonymisedPreviousStudy}. Although our categorisation was too granular to compare with these, the debugging strategies students used likely affected and were affected by the emotions they experienced in the debugging exercises. For example, behaviours in line with tinkering or extreme moving may have expressed an emotional disconnect with the exercises \citep{ConditionsLearningNovices} or a lack of motivation to complete the debugging exercises.

There has been little research comparing students' debugging behaviours with their attitudes and emotions. Previous research on the emotional aspect of learning to program has found encountering and struggling with errors to be particularly distressing moments \citep{EDAProgrammingEmotions, ProgrammingAssignmentsEmotionalToll} which can negatively contribute to students' programming self-efficacy \citep{ProgrammingSelfEfficacySelfAssessment, ProgrammingSelfEfficacySelfAssessment}. Further investigating how these moments are associated with debugging behaviour, and providing support for emotional regulation in beginner programming environments, is important work to conduct.

\section{Limitations and Reflections on Analysing Debugging Behaviour}\label{sec:limitations}
There are several limitations with our study design and content analysis that affect the generalisability, reliability, and validity of our findings. Regarding the study, the sample of students was male-dominated and taken from two relatively high-performing schools. Additionally, we did not collect students' programming experience or familiarity with debugging strategies, meaning some students may have spent extensive time programming outside of the classroom. Like many similar studies \citep{InvestigatingDebuggingProcessesScopingReview}, the use of foreign code meant that the behaviours we captured were not representative of how students debug their own programs. While students will certainly debug foreign programs at points, analysing students' debugging behaviour `in the wild' is important for understanding students' classroom experiences with learning to program. Future work could do this by collecting students' in-class programming sessions and analysing debugging episodes \citep{TowardsTheoryBasedAnalysisDebuggingProcesses}, or by getting students to first complete some programming exercises that are closely related to later debugging exercises \citep{FindingFixingFlailing}.

As discussed in Section \ref{sec:difficulty-agreeing-categorisation}, the lack of agreement in our qualitative content analysis meant we could not reliably compare the frequency of debugging behaviours as we originally intended. Even if we had achieved higher levels of agreement, we could have only expressed the prevalence of behaviours in terms of exercise attempts---students who reverted one or ten changes would have been coded with the same behaviour. Our content analysis also ignored resolutions of errors that students had introduced themselves, which caused us to miss certain behaviours. Finally, the majority of the behaviours in the categorisation focused on between-run changes rather than higher-level behaviours, which prevented comparison with students' attitudes and emotions towards debugging.

These decisions were ultimately made to balance the contribution of our results with the soundness of our analysis. Manually replaying program snapshots can reveal rich insights into students' programming process \citep{JadudCompilationBehaviour, WritingBetweenLines} that cannot be found through quantitative analysis or examination of single snapshots. In addition to exploratory research, program snapshot analysis can be used to provide feedback to teachers and students and evaluate the effect of pedagogical interventions on behaviour. However, qualitatively coding based on this manual inspection is difficult. In this study, we primarily coded between-run changes per exercise attempt to avoid unjustified interpretation and the difficulty of unitising debugging behaviours \citep{WritingBetweenLines}. However, this yielded a large categorisation that was difficult to consistently apply and hence compare with other studies. Similarly, coding high-level strategies is near-impossible to reliably perform with program snapshot data alone due to the intentions and physical components involved in strategy use.

This introduces a discrepancy between the utility of manually inspecting programming process data and the difficulty of analysing it, which other researchers have noted. \citet{WritingBetweenLines}, for example, thematically analysed novice programmers' line-by-line edits by replaying over 100 programming sessions. When reflecting on their analysis, they similarly conclude that both ``attempting to tag all activity is fruitless'' and ``tagging long, higher-level behaviours was hard to do reliably'' \citep[p. 168]{WritingBetweenLines}.

How can this useful method of investigating students' programming process be reliably conducted? Based on our experience, we believe the best approach is to triangulate snapshot data with student input, such as retrospective interviews or think-aloud protocols. This would help to understand the strategic intent behind students' actions. For larger datasets, another effective approach would be combining manual inspection of snapshot data with quantitative analysis. This could be done similarly to our study, where our quantitative analysis supplemented our qualitative findings, or quantitative analysis could be combined with in-depth analysis of some representative case studies. Either way, researchers who conduct similar future research should consider combining such methods and carefully consider what claims their analysis allows them to make.


\section{Conclusions and Future Work}\label{sec:conclusions-further-work}
This study has investigated the debugging behaviour of secondary school students learning to program in Python. By manually replaying program snapshots, we performed a qualitative content analysis of 73 students' attempts at a set of debugging exercises that yielded a granular categorisation of students' debugging behaviour.

The categorisation highlighted a range of ineffective and effective behaviours that students use to debug foreign Python programs. Students frequently attempted early fixes before giving themselves time to understand the actual or intended behaviour of the program, reverted corrective changes, added errors, and repeatedly ran identical programs in quick succession. Small-scale and quick-fire changes were also common, reminiscent of the tinkering and extreme moving \citep{ConditionsLearningNovices} that has been observed in previous studies of school students' debugging behaviour \citep{ProblemSolvingDebuggingK68, ElementaryPuzzleBasedDebugging, MAADSMethod, ElementaryGameBasedDebugging}. When students did resolve errors, their changes often appeared more targeted, though even some corrective changes introduced additional errors.

Our findings led us to consider barriers that may be preventing students from reliably and successfully resolving errors. Students may have fragile knowledge of the programming constructs they are fixing \citep{ComparisonDebuggingBehaviourNoviceExpert, NottinghamDebuggingBehaviour, ThinkAloudNoviceDebugging}, lack a systematic approach that would foster more targeted and intentional changes, struggle with the syntax barrier in Python \citep{UnderstandingSyntaxBarrier}, or have attitudes and experience emotions that impede their ability to successfully debug. To help alleviate these barriers, our work has the following implications for school teachers wishing to improve their debugging teaching practice.

\begin{itemize}
    \item To improve the systematicity of students' debugging behaviours, teachers could explicitly model a systematic debugging process, such as those already developed by researchers \citep{CarverImprovingChildrensDebugging, SystematicProcessMichaeli, ExplicitlyTeachingDebuggingPrimarySchool}. Encouraging students to complete the initial stages of debugging before attempting a fix is particularly important. 
    \item Ineffective debugging behaviours, such as quick cycles of small and near-random changes, should be discouraged. This could be done by modelling the consequences of employing such behaviours, which could then motivate the use of more systematic strategies.
    \item Students could use a programming environment that alleviates the common struggles they have with the syntax of text-based programming languages, such as frame-based environments (e.g., \citep{FrameBasedEditing, StrypePoster2024, TeachingCodingDifferentlyElan}).
\end{itemize}

We believe these suggestions are particularly well-suited to school teachers who are not experienced or confident programmers. Unlike specialist debugging tools, modelling the debugging process or teaching with a different programming environment does not require programming expertise, especially as support for these has already exists.

Our study is one of the first in the context of school students' debugging behaviour in a TBPL. Given the increasing number of curricula mandating the teaching of TBPLs in secondary schools (e.g., \citep{UKSecondaryComputingCurriculum, ScotlandTechnologiesCurriculum, Swedish2024Curriculum, BritishColumbiaADSTCurriculum, ProgrammingSkillsInLatinAmerica}), further research that investigates and improves students' debugging behaviour could benefit the millions of students who are learning to program. Therefore, we suggest the future work for researchers and developers of beginner programming environments:

\begin{itemize}
    \item Future studies of students' debugging behaviour should combine their \textit{actual} behaviours, through methods such as logging or screen recordings, and \textit{reported} behaviours, through methods such as retrospective interviews or think-aloud protocols. If manual inspection is used to analyse students' actual behaviours, it should be combined with quantitative analysis. This would give a clearer idea of students' intentional debugging strategies, which we were not able to conclude.
    \item Designers of beginner programming environments should incorporate lightweight features that encourage a systematic debugging approach, and research the effect of these on debugging behaviour.
    \item Researchers should investigate the long-term effect of pedagogical approaches for debugging on students' debugging behaviour to understand whether students adopt the approaches they have been taught.
\end{itemize}

This research will help to improve students' debugging ability and their experiences with learning to program.


\balance

\bibliographystyle{ACM-Reference-Format}
\bibliography{bibliography.bib, anonymisedbib}

\appendix
\section{Full Categorisation of Students' Debugging Behaviour}\label{appendix-a}
\begin{longtable}{p{6.35cm} c c}
\label{tab:category-book} \\
        \toprule
        \textbf{Category} & \textbf{Student count (\begin{math}n=73\end{math})} & \textbf{Exercise attempt count (\begin{math}n=322\end{math})} \\
        \midrule
        \endfirsthead
        \toprule
        \textbf{Category} & \textbf{Student count (\begin{math}n=73\end{math})} & \textbf{Exercise attempt count (\begin{math}n=322\end{math})} \\
        \midrule
        \endhead
        \textbf{1. First move} & & \\
        \hspace{0.5cm} Made changes before running program & 47 (65\%) & 110 (34\%) \\
        \hspace{0.5cm} Ran program before making changes & 69 (95\%) & 212 (66\%) \\
        
        \midrule

        \textbf{2. Positive debugging behaviours} & 71 (97\%) & 259 (80\%) \\
        \hspace{0.5cm} Entered incorrect input & 3 (4\%) & 3 (1\%) \\
        \hspace{0.5cm} Made improvements to program & 2 (3\%) & 5 (2\%) \\
        \hspace{0.5cm} First change on line referred to in error \\ \hspace{0.5cm} message & 71 (97\%) & 257 (80\%) \\

        \midrule
         
        \textbf{3. Added errors} & 69 (95\%) & 245 (76\%) \\
        \hspace{0.5cm} Reverted corrective change(s) & 33 (45\%) & 43 (13\%) \\
        \hspace{0.5cm} Added \textbf{syntax} error(s) & 63 (86\%) & 184 (57\%) \\
        \hspace{1cm} Changes to \textbf{existing statements} & 59 (81\%) & 158 (49\%) \\
        \hspace{1.5cm} Changes to \textbf{output statements} & 42 (58\%) & 67 (21\%) \\
        \hspace{1.5cm} Changes to \textbf{selection} & 39 (53\%) & 51 (16\%) \\
        \hspace{1.5cm} Changes to \textbf{iteration} & 18 (25\%) & 18 (6\%) \\
        \hspace{1.5cm} Changes to \textbf{variable assignments} & 41 (56\%) & 66 (20\%) \\
        \hspace{1.5cm} Changes to \textbf{other statements in} \\ \hspace{1.5cm} \textbf{program} & 3 (4\%) & 3 (1\%) \\
        \hspace{1cm} Changes involving \textbf{symbols} & 48 (66\%) & 92 (29\%)\\
        \hspace{1.5cm} Changes to \textbf{operators} & 25 (34\%) & 34 (11\%) \\
        \hspace{2cm} Mathematical operators & 25 (34\%)& 34 (11\%) \\
        \hspace{2cm} Logical operators & 2 (3\%) & 2 (1\%) \\
        \hspace{1.5cm} Changes to \textbf{other symbols} & 38 (52\%) & 64 (20\%) \\
        
        \hspace{1cm} Changes involving fusion or distribution \\ \hspace{1cm} of \textbf{multiple lines of code} & 21 (29\%) & 27 (8\%) \\
        \hspace{1cm} Addition of \textbf{non-Python strings} & 14 (19\%) & 19 (6\%) \\
        \hspace{1cm} Changes involving \textbf{variable references} & 15 (21\%) & 17 (5\%) \\
        \hspace{1cm} Changes to \textbf{whitespace of program} & 28 (38\%) & 45 (14\%) \\

        \hspace{0.5cm} Added \textbf{runtime} error(s) & 39 (53\%) & 61 (19\%) \\
        \hspace{1cm} Changes to \textbf{variable assignments} & 31 (42\%) & 45 (14\%) \\
        \hspace{1cm} Changes to \textbf{variable references} & 12 (16\%) & 16 (5\%) \\
        \hspace{1cm} Changes to \textbf{other statements in} \\ \hspace{1cm} \textbf{program} & 10 (14\%) & 10 (3\%) \\

        \hspace{0.5cm} Added \textbf{type} error(s) & 33 (45\%) & 45 (14\%) \\
        \hspace{1cm} Changes involving \textbf{variables} & 24 (33\%) & 27 (8\%) \\
        \hspace{1cm} Changes to \textbf{function call} & 12 (16\%) & 15 (5\%) \\
        \hspace{1cm} Changes to \textbf{other statements} \\ \hspace{1cm} \textbf{program} & 15 (21\%) & 17 (5\%) \\

        \hspace{0.5cm} Added \textbf{logical} error(s) & 53 (73\%) & 94 (29\%) \\
        \hspace{1cm} Changes involving \textbf{output statements} & 12 (16\%) & 14 (4\%) \\
        \hspace{1cm} Changes to \textbf{operators} & 21 (29\%) & 26 (8\%) \\
        \hspace{1.5cm} Mathematical operators & 16 (22\%) & 18 (6\%) \\
        \hspace{1.5cm} Logical operators & 8 (11\%) & 8 (2\%) \\
        \hspace{1cm} Changes involving \textbf{variables} & 26 (36\%) & 34 (11\%) \\
        \hspace{1cm} Changes to \textbf{program flow} & 29 (40\%) & 38 (12\%) \\
        \hspace{1cm} Changes to \textbf{other statements} & 3 (4\%) & 3 (1\%) \\

        \midrule
         
        \textbf{4. Resolved errors} & 66 (90\%) & 268 (83\%) \\

        \hspace{0.5cm} Resolved \textbf{syntax} error(s) & 65 (89\%) & 206 (64\%)\\
        \hspace{1cm} Correctly resolved syntax error & 59 (81\%) & 148 (46\%) \\
        \hspace{1cm} Resolved syntax error by changing syntax \\ \hspace{1cm} of erroneous component & 18 (25\%) & 23 (7\%) \\
        \hspace{1cm} Introduced logical error from resolution \\ \hspace{1cm} of syntax error & 39 (53\%) & 46 (14\%) \\
        \hspace{1cm} Hard-coded resolution of syntax error & 5 (7\%) & 5 (2\%) \\

        \hspace{0.5cm} Resolved \textbf{runtime} error(s) & 61 (84\%) & 100 (31\%) \\
        \hspace{1cm} Correctly resolved runtime error & 43 (59\%) & 67 (21\%) \\
        \hspace{1cm} Introduced logical error from resolution \\ \hspace{1cm} of runtime error & 14 (19\%) & 16 (5\%) \\
        \hspace{1cm} Hard-coded resolution of runtime error & 17 (23\%) & 17 (5\%) \\

        \hspace{0.5cm} Resolved \textbf{type} error(s) & 34 (47\%) & 34 (11\%) \\
        \hspace{1cm} Correctly resolved type error & 26 (36\%) & 26 (8\%) \\
        \hspace{1cm} Hard-coded resolution of type error & 8 (11\%) & 8 (2\%) \\

        \hspace{0.5cm} Resolved \textbf{logical} error(s) & 56 (77\%) & 170 (53\%) \\
        \hspace{1cm} Correctly resolved logical error & 55 (75\%) & 164 (51\%) \\
        \hspace{1.5cm} While code not running & 53 (73\%) & 103 (32\%) \\
        \hspace{1.5cm} After a single successful execution & 27 (37\%) & 42 (13\%) \\
        \hspace{1.5cm} After repeated successful executions & 35 (48\%) & 63 (20\%) \\
        \hspace{1cm} Hard-coded resolution of logical error & 19 (26\%) & 21 (7\%) \\

        \midrule
        \textbf{5. Inconsequential changes} & 71 (97\%) & 209 (65\%) \\
        \hspace{0.5cm} Added, removed, or edited of \textbf{comments} & 7 (10\%) & 9 (3\%) \\
        \hspace{0.5cm} Changes involving \textbf{symbols} & 30 (41\%) & 52 (16\%) \\
        \hspace{0.5cm} Changes involving \textbf{variable references} & 4 (5\%) & 4 (1\%) \\
        \hspace{0.5cm} Changes to \textbf{whitespace of program} & 64 (88\%) & 136 (42\%) \\
        \hspace{0.5cm} Changes to \textbf{existing statements in the} \\ \hspace{0.5cm} \textbf{program} & 66 (90\%) & 152 (47\%) \\
        \hspace{1cm} Outputs & 48 (66\%) & 69 (21\%) \\
        \hspace{1cm} Inputs & 19 (26\%) & 22 (7\%) \\
        \hspace{1cm} Other statements & 55 (75\%) & 91 (28\%) \\
        \hspace{0.5cm} Addition of \textbf{unnecessary code} & 15 (21\%) & 20 (6\%) \\
        
        \midrule
        \textbf{6. Miscellaneous behaviours} & 70 (96\%) & 243 (75\%) \\
        \hspace{0.5cm} Repeated runs & 69 (95\%) & 205 (64\%) \\
        \hspace{1cm} Repeated ran successfully executing \\ \hspace{1cm} program & 49 (67\%) & 89 (28\%) \\
        \hspace{1cm} Repeatedly ran erroneous program & 55 (75\%) & 135 (42\%) \\
        \hspace{1cm} Repeatedly ran program at \textbf{beginning} of \\ \hspace{1cm} exercise & 15 (21\%) & 16 (5\%) \\
        \hspace{0.5cm} Reverted previous changes & 61 (84\%) & 159 (49\%) \\
        \bottomrule
\end{longtable}
\section{Debugging Exercises}\label{appendix-b}
Each exercise contains details of the original description and program presented to students, as well as a table detailing the exact errors and how they could be resolved. The categorisation of syntax, semantic, and type errors is based on the error message raised by the code editor used for the study. The erroneous lines of the program, or specific constructs where appropriate, are highlighted in yellow.

\subsection{Exercise 1}
Alex has measured out the dimensions of a cuboid to use for a science experiment. Alex has made a program to calculate the volume of the cuboid and print it to the screen. However, the program contains some errors.

The height and depth of the cuboid are 25cm and its width is 50cm. The volume of a cuboid is calculated using the equation below:
\begin{itemize}
    \item Volume = height x width x depth
\end{itemize}
This program contains 2 errors - have a go at fixing them all.

\begin{lstlisting}[style=studentsnapshotlisting, title=The original program for exercise 1. Error message: \texttt{SyntaxError: bad input on line 6}, escapechar=!]
# Question 1
!\colorbox{yellow}{height = depth}!
width = 50
depth = 25
volume = height * width * depth
!\colorbox{yellow}{print("The volume is,volume)}!
\end{lstlisting}

\begin{table}[h]
    \caption{The errors in exercise 1}
    \label{tab:exercise-1-error-details}
    \begin{tabular}{c c p{5.25cm} p{4.25cm}}
        \toprule
        \textbf{Erroneous line} & \textbf{Type of error} & \textbf{Description} & \textbf{Example fix} \\
        \midrule
        Line 2 & Semantic & The \texttt{depth} variable is being referred to before it has been assigned a value. & Move the assignment to after the assignment on line 4. \\
        Line 4 & Syntax & The program is missing a closing quotation mark before the closing paranthesis. & Add a quotation mark (\texttt{"}) between \texttt{volume} and \texttt{)} on line 4. \\
        \bottomrule
    \end{tabular}
\end{table}

\subsection{Exercise 2}
This program inputs the user's first name, surname, and the year they were born. It then prints a sentence to the screen with their full name and how old they will be at the end of the year.

If a user's first name is Jo, their last name is Bloggs, and they were born in 2008, the program should print: "Your name is Jo Bloggs and at the end of this year you will be 15".

This program has 3 errors - have a go at fixing them all.

\begin{lstlisting}[style=studentsnapshotlisting, title=The original program for exercise 2. Error message: \texttt{SyntaxError: bad input on line 3}, escapechar=!]
# Question 2
!\colorbox{yellow}{input("What year were you born in? ") = year\_born}!
!\colorbox{yellow}{ }!age = 2023-int(year_born)

first_name = input("What is your first name? ")
last_name = input("What is your last name? ")
print("Your name is",first_name,last_name,"and at the end of this year you will be !\colorbox{yellow}{age"}!)
\end{lstlisting}

\begin{table}[h]
    \caption{The errors in exercise 2}
    \label{tab:exercise-2-error-details}
    \begin{tabular}{c c p{5cm} p{4.5cm}}
        \toprule
        \textbf{Erroneous line} & \textbf{Type of error} & \textbf{Description} & \textbf{Example fix} \\
        \midrule
        Line 1 & Syntax & The expressions in the assignment are the wrong way around. & Switch around the terms in the assignment in line 1. \\
        Line 2 & Syntax & There is an invalid indentation at the beginning of the line. & Remove the indentation from line 2. \\
        Line 7 & Logical & The string \texttt{age} is being printed rather than the value of the variable. & Reposition the final quotation mark on line 7 to be before \texttt{age} and proceed with a comma (i.e., \texttt{", age)}. \\
        \bottomrule
    \end{tabular}
\end{table}

\subsection{Exercise 3}
This program checks if someone should apply to be a computing teacher using the steps below:
\begin{itemize}
    \item Input the user's age.
    \item Input the user's response to the question "Do you have a passion for teaching computing? Enter 'yes' or 'no': "
    \item If the user is 21 or over and does have a passion for teaching computing, the check should be a success. Otherwise, the check should be unsuccessful.
    \item Print the result of the check.
\end{itemize}
This program has 4 errors - have a go at fixing them all.

\begin{lstlisting}[style=studentsnapshotlisting, title=The original program for exercise 3. Error message: \texttt{SyntaxError: bad input on line 6}, escapechar=!]
# Question 3
print("This program will check if you should apply to be a computing teacher")
age = int(input("What is your age? "))
computing_degree = input("Do you have a passion for teaching computing? Enter 'yes' or 'no': ")

if age !\colorbox{yellow}{>}! 21 !\colorbox{yellow}{or}! computing_degree !\colorbox{yellow}{=}! "yes":
  allowed_to_apply = "Successful"
else:
  allowed_to_apply = "Unsuccessful"
!\colorbox{yellow}{ }!print("Result of check:",allowed_to_apply)
\end{lstlisting}

\begin{table}[h]
    \caption{The errors in exercise 3}
    \label{tab:exercise-3-error-details}
    \begin{tabular}{c c p{5.25cm} p{4.25cm}}
        \toprule
        \textbf{Erroneous line} & \textbf{Type of error} & \textbf{Description} & \textbf{Example fix} \\
        \midrule
        Line 6 & Syntax & The assignment operator (\texttt{=}) is used in a syntactically invalid position. & Replace the assignment operator (\texttt{=}) with the equality operator (\texttt{==}) on line 6 . \\
        Line 6 & Logical & Incorrect inequality operator (\texttt{>}) causing the program to output the incorrect statement when \texttt{age = 21}. & Replace the greater than operator (\texttt{>}) on line 6 with a greater than or equal operator (\texttt{>=}). \\
        Line 6 & Logical & Incorrect logical operator (\texttt{or}) causes the program to output the incorrect error message when only one of the conditions in the if statement is true. & Replace the \texttt{or} operator on line 6 with an \texttt{and} operator.  \\
        Line 10 & Logical & The line is indented, meaning the statement is only printed when \texttt{} instead of always being printed. & Unindent line 10. \\
        \bottomrule
    \end{tabular}
\end{table}

\subsection{Exercise 4}\label{sec:appendix-exercise-4}
This program should display the first 12 multiples of the number 6 using a while loop, from 1x6 up to 12x6. Fix the errors to make sure it prints them properly.

This program has 4 errors - have a go at fixing them all.

\begin{lstlisting}[style=studentsnapshotlisting, title=The original program for exercise 4. Error message: \texttt{TypeError: \textquotesingle <\textquotesingle \  not supported between instances of \textquotesingle str\textquotesingle \ and \textquotesingle int\textquotesingle \  on line 4} (after \texttt{"The first 12 multiples of the number 6 are:"} is printed), escapechar=!]
# Question 4
count=!\colorbox{yellow}{"1"}!
print("The first 12 multiples of the number 6 are:")
while count!\colorbox{yellow}{<}!12:
  times_table = 6 * count
  count=count+1
!\colorbox{yellow}{print(times\_Table)}!
\end{lstlisting}

\begin{table}[h]
    \caption{The errors in exercise 4}
    \label{tab:exercise-4-error-details}
    \begin{tabular}{c c p{5.25cm} p{4.25cm}}
        \toprule
        \textbf{Erroneous line} & \textbf{Type of error} & \textbf{Description} & \textbf{Example fix} \\
        \midrule
        Line 2 & Type & The \texttt{count} variable is a string rather than an integer, causing a type error on line 4. & Remove quotation marks from \texttt{1} on line 2. \\
        Line 4 & Logical & The iteration logic for this program causes the first 11 multiples of 6 to be printed rather than the first 12. & Replace the less than operator (\texttt{<}) on line 4 with a less than or equal operator (\texttt{<=}). \\
        Line 7 & Semantic & The variable \texttt{times\_Table} does not exist in the program. & Change spelling of \texttt{times\_Table} on line 7 to \texttt{times\_table}. \\
        Line 7 & Logical & This print statement is only executed at the end of the program rather than for each multiple. & Indent line 7 so that it becomes a part of the while loop logic. \\
        \bottomrule
    \end{tabular}
\end{table}

\subsection{Exercise 5}
This program simulates a game of rock, paper, scissors between two players. This is done using the following steps:

\begin{itemize}
    \item Randomly generate a choice for each player using the random.choice() function, which selects a random item from a given list.
    \item Work out which player won the game or if it was a draw.
    \item Print the winner to the screen. If the match is a draw, then "The winner is Neither" should be printed.
\end{itemize}

This program has 4 errors - have a go at fixing them all.

\begin{lstlisting}[style=studentsnapshotlisting, title=The original program for exercise 5. Error message: \texttt{SyntaxError: bad input on line 16}, escapechar=!]
# Question 5
import random
player_1 = random.choice(["Rock","Paper","Scissors"])
print("Player 1 has chosen "+player_1)
player_2 = random.choice(["Rock","Paper","Scissors"])
print("Player 2 has chosen "+player_2)

if player_1 == player_2:
  winner = "Neither"
elif player_1 == "Rock" and player_2 == "Scissors":
  winner = "Player 1"
elif player_1 =="Paper" !\colorbox{yellow}{or}! player_2 == "Rock":
  winner = "Player 1"
elif player_1 == "Scissors" !\colorbox{yellow}{or}! player_2 == "Paper":
  winner = "Player 1"
!\colorbox{yellow}{elif}!:
  winner = "Player 2"
!\colorbox{yellow}{ }!print("The winner is "+winner)
\end{lstlisting}

\begin{table}[h]
    \caption{The errors in exercise 5}
    \label{tab:exercise-5-error-details}
    \begin{tabular}{c c p{5.2cm} p{4cm}}
        \toprule
        \textbf{Erroneous line} & \textbf{Type of error} & \textbf{Description} & \textbf{Example fix} \\
        \midrule
        Line 12 & Logical & Assigns the winner to player 1 if player 1 is paper \textit{or} player 2 is rock, allowing player 1 to be assigned as the winner in incorrect conditions (e.g., if player 1 is paper and player 2 is scissors). & Replace the \texttt{or} operator on line 12 with an \texttt{and} operator. \\
        Line 14 & Logical & Assigns the winner to player 1 if player 1 is scissors \textit{or} player 2 is paper, allowing player 1 to be assigned as the winner in incorrect conditions (e.g., if player 1 is rock and player 2 is paper) & Replace the \texttt{or} operator on line 14 with an \texttt{and} operator. \\
        Line 16 & Syntax & \texttt{elif} is not followed by any condition, so is syntactically invalid. & Replace \texttt{elif} with \texttt{else}. \\
        Line 18 & Logical & This print statement should always be executed but is currently not. & Unindent the line. \\
        \bottomrule
    \end{tabular}
\end{table}

\end{document}